# A possible mechanism for the negative capacitance observed in organic devices

X. Q. Wang and C. B. Cai

Physics Department, Shanghai University, Shanghai 200444, China

**Abstract:**

The mechanism of negative capacitance, e.g. inductance, induced by a sufficient electrical field in the organic device is investigated. The cations in organic bulk are proposed to be driven by the applied voltage and to accumulate at the interface, and further to generate the surface states or media states. These states result in a larger junction current through the device, indicating the negative capacitances which are simulated in three situations: impedance spectrum, capacitance measurement and current response. This simple kinetic model may be helpful to understand why the negative capacitance phenomenon is observed in various organic devices.

Corresponding author' mail:

shawl.wang@gmail.com (Xiaoqi Wang)

cbcai@shu.edu.cn (Chuanbing Cai)

During the past decades, the negative capacitance (NC) phenomenon observed in many organic heterostructure devices, such as organic solar cells, organic electronics and optoelectronic devices, has been attracted great interests[1-3]. The NC behavior can be identified by several measurements, including a lower semicircle in the electric impedance measurements (EIS)[4], a gradually transition of device capacitance from positive value to negative recorded by LCR meter[5], and an increasing current response after applying a voltage step in dc measurements[6], etc.

To understand the origination of NC behavior, many possible mechanisms have been proposed for various devices and most of them are concluded by H. Kliem[7]. These proposals summary this phenomenon into three aspects: a) the negative capacitance is subjected to the junction interface and b) can be driven by the external force, such as voltage, and c) Whatever physical origination responsible for NC phenomenon it must be temporal evolution[7]. For organic devices, however, it is worth noting that the cations will be introduced into the organic bulk by intentionally[8] or unintentionally[9]. These small size cations, i.e. $Na^+$ and $Al^+$, can be driven by an external electric field, and gradually accumulate at the inorganic/organic interface[8, 9]. This kinetic process has also been adopted in explanation for the resistive switching behavior of inorganic/organic heterojunctions, in which the accumulated cations are assumed to produce a thick and high enough barrier to promote electron transfer through the junction through tunneling, resulting in a switching in resistance[8, 9]. In this paper, we extend the idea and suggest that the accumulation of cations at interface may generate several intermediate states or surface states, allowing a larger current through the junction. Furthermore, the generation of states is assumed to depend on the external electric field, which is responsible for the NC behavior observed in measurement.

In the simple kinetic model, our discussions focus on the interface of heterojunction and the cation dynamics in organic bulk, rather than the inorganic part no matter it is semiconducting or insulating. Figure 1(a) models a virtual heterojunction consists of inorganic semiconductor and organic materials. Assumed that by applying a voltage the cations dispersed in organic bulk will be driven and locally accumulate at the inorganic/organic interface, which generates several intermediate states or surface states within the depletion layer, as indicated in Fig. 1(b) and (c). With the temporal evolution, more and more cations drift into the depletion layer and cover on the cross section of the junction, increasing the proportion $\theta$ of the covered area. The change rate of $\theta$ can be approximately considered to has a first-order relationship $d\theta/dt = k_f(1-\theta) - k_b\theta$, where $k_f$ is the generation rate of intermediate states and $k_b$ the destruction rate. In organic bulk, the movement of cations is through hopping from one lattice site to another. The hopping frequency $v$ can be approximately described by the Vogel-Tammann-Fulcher (VTF) relationship[10] $v = v_0 \exp[q(V_a - V_{act})/k_B(T - T_0)]$, where $v_0$ is the hopping frequency constant, $V_{act}$ the activation potential, $T_0$ the reference temperature at which the hopping of cation goes to zero, $q$, $k_B$, and $T$ are the elementary charge, Boltzmann constant and temperature, respectively. Cations hopping to the interface will generate intermediate states while they hopping back into the organic bulk will destruct these states. Therefore, $k_f$ and $k_b$ are considered to be proportional to $v$ with an exponential relationship of $V$, and have a similar exponential expression as $k_f = k_0 \exp[\gamma(V_a - V_{eq})]$ and $k_b = k_0 \exp[-\gamma(V_a - V_{eq})]$, respectively, where $k_0$ is constant coefficient, $V_{eq}$ the equilibrium potential, at which the generation of states is counterbalance to the destruction, and $\gamma$ the parameters containing $q$, $k_B$, $T$, $T_0$.

These generated states are advantageous for electron transfer through the junction by electron injection from the inorganic states first to the generated intermediate states in depletion layer and then to the organic bulk states[11]. It results in an additional larger transmission current density $J_T$ in the region covered with cations, relative to the virginal heterojunction current density $J_D$ in other uncovered regions, seen in Fig. 1(b). As the device in discussion is a Schottky type, the virginal current flowing through the junction can be described as a diode type as $J_D = J_d[\exp(\beta V_a) - 1]$, where $J_d$ is the exchange current density, and $\beta = q/nk_BT$ with a ideality factor $n$. Due to the electron transfer through the intermediate states happens in the identical device, it is assumed that $J_T$ has a similar expression to $J_D$ and can be written as $J_T = J_t[\exp(\alpha V_a) - 1]$, where $J_t$ and $\alpha$ have the same physical meaning as $J_d$ and $\beta$. Therefore, the junction current density through the junction device can be deduced as $J_p = \theta \cdot J_T + (1-\theta) \cdot J_D$.

During the EIS measurement, a small periodic perturbation voltage signal $\Delta V$ superposition to the external voltage $V$ is applied to the device. The measured impedance of device can be described by $Z_p = Y^{-1} = (\Delta J / \Delta V_a)^{-1}$, where $\Delta J$ denote a small signal of response current density. Within the steady-state regime[12], the impedance $Z$ is obtained through linearizing the total current density with respect to $\Delta V$ and $\Delta\theta$ and subsequently solved for $Z_p$ by eliminating $\Delta\theta$, as seen eq. (1).

$$Z_p = Y^{-1} = \left(\frac{dJ}{dV_a}\right)^{-1} = \left(\frac{1}{R_p} + \frac{1}{jwL + R_s}\right)^{-1} \tag{1}$$

Where, $R_p = [(\alpha J_T - \beta J_D)\theta + \beta J_D]^{-1}$, $R_s = \{\gamma\tau(J_T - J_D)[k_f(1-\theta) + k_b\theta]\}^{-1}$, $L = \{\gamma(J_T - J_D)[k_f(1-\theta) + k_b\theta]\}^{-1}$, and $\theta = k_f\tau[1 - \exp(-t/\tau)]$ with $\tau = (k_f + k_b)^{-1}$. The calculated impedance suggests a parallel circuit of $R_p$ and $R_s$ series with an equivalent inductor $L$ which represents a negative capacitance[7]. In physically, $R_p$ denotes the dc differential resistance of the junction, while $R_s$ and $L$ represent the dynamical processes induced by the small periodic perturbation. These equivalent elements are all subjected to the applied voltage $V$ and the proportion $\theta$, implying they are temporal evolution.

For a realistic heterojunction device, there are a barrier capacitance in the built-in region and a diffusion capacitance in the depletion layer. Both of them, in this paper, are represented by a constant junction capacitance $C$ for the simplicity but without losing the physical meaning. Furthermore, the bulk resistance $R_b$ for inorganic and organic materials is also taken into account. Different from a tunneling diode[13], the junction capacitance $C$ is considered to be parallel connected with the inductive part $Z_p$ due to the simultaneously occurrence of kinetics of cations and the effect of junction capacitance, as illustrated in the Figure 2(a). Noted that IS measurements are generally achieved under the steady state, which means the generation and destruction of the intermediate states are counterbalance, i.e. $d\theta/dt = 0$. The total impedance $Z$ is then derived by $R_b$ series with a parallel circuit of $Z_p$ and $C$.

Figure 2(b) indicates that the Nyquist plot of the calculated $Z$ under the steady state. The impedance curves are simulated during the frequency from $10^5$Hz to 0.1Hz. The values of other parameters are listed in the caption of Figure 3. The intermediate states will be significantly generated as the applied voltage larger than the equilibrium potential which is set as $V_{eq}$=1V. In the plot there is an upper semicircle can be observed, which is subjected to the junction capacitance. With the applied voltage is increasing, the semicircle becomes smaller and $Z$" varies into the lower plane, suggesting the negative capacitive behavior. For a clear illustration, the impedance curve in the rectangle frame is enlarged and replotted in Fig. 2 (c). As $V_a$>$V_{eq}$, cations in organic bulk will be driven towards the interface with the increasing applied voltage and will diffuse back to the organic bulk as the voltage is decreasing. The proportion $\theta$ of the generated intermediate states will correspondingly vary with the variation of the applied voltage, which is responsible for the lower semicircles observed in the fourth quadrant in dynamical measurements.

To clarify the voltage dependence of transition from a positive capacitance to a negative one, the frequency dependence of both the total capacitance $C_{tot}$ and the dissipation factor tan δ are illustrated in the Figure 3. The measurements simulated here are usually achieved by LCR meter[5]. In a parallel measure mode[14], the device is viewed as an RC parallel circuit with admittance $Y = 1/R + jwC_{tot}$. In our simulation, $C_{tot}$ is derived from the imaginary admittance $Y$" and tan δ is calculated by quality factor[14]. Under a sufficient voltage $V_a$, there is a gradually transition from positive value to negative one in the $C_{tot}$ versus. frequency (left axis), and correspondingly a sharp peak at the position where $C_{tot}$ is zero in curves of tan δ (right axis). As $V_a$ is enhanced, the peak shifts towards high frequency region, suggesting a voltage dependence of the negative capacitance behavior as reported in Ref. 7. It can be understood by the present kinetic model because in which the proportion $\theta$ will increase as the enhanced $V_a$, which leads to $Y_p$ decreasing and further affects the total capacitance $C_{tot}$ represented by $Y_p$ and $C$ in mathematically.

Figure 4 shows the transient response of the device current $J_{tot}$ evaluated by a self-consistent calculation complementary to the finite-difference method, where $J_{tot} = \theta \cdot J_T + (1-\theta) \cdot J_D - R_b C \cdot dJ_{tot}/dt$. In the Fig. 4(a), the response current indicates a capacitance dependence decay as $V_a$ is 0.5 V, but behaves an inductor dependent increasing as the $V_a > V_{eq}$, as the expectation in the present model. Fig. 4(b) demonstrates the NC behavior is dependent on cation accumulation rate $k_0$ the larger $k_0$ the faster $I_{tot}$ increasing. These similar phenomenons are also observed in the device of Al/P(VDF-TrFe)/siO2/nSi[7]. Out of the cation accumulation regime, i.e. $k_0=0$, the response current decay is only subjected to the junction capacitance.

In conclusion, we propose a kinetic model, in which cations dispersed in organic bulk will be driven by the external electric field and locally accumulate at the inorganic/organic interface. The intermediate states or surface states are thus generated that are advantageous for electron transfer through the junction. The kinetic model denotes the dynamics of generation and destruction of intermediate states should be responsible for the negative capacitance observed in various measurements. For the realistic device the dynamical process will be complex and the description of the generation and destruction rate of intermediate states may need to be revised. But we note this simple model will still be the basic viewpoint which is helpful to understand why the NC phenomenon can be observed in various organic devices.

**Acknowledgement:**

This work is partly sponsored by the Innovation Funds for Ph.D. Graduates of Shanghai University (2010), China.

**FIGURE CAPTIONS:**

Fig. 1.

Schematic diagrams for the inorganic/organic device: (a) Without the cation accumulation at the interface, (b) with the cation accumulation. Arrow denotes the direction of electron flow, and (c) energy level diagram for the inhomogeneous interface.

Fig. 2

Simulation for the impedance spectrum: (a) the equivalent electrical circuit calculated by the present kinetic model. (b) Nyquist plot of Z" vs. Z'. The part of plot in the rectangle frame is enlarged and replotted in (c) for clarity. The parameters in simulation are listed: $J_t=10^{-5}$ A/cm$^2$; $J_d=10^{-7}$ A/cm$^2$; $k_0=1$ s$^{-1}$; $\alpha=2$ V$^{-1}$; $\beta=1$ V$^{-1}$; $\gamma=1$ V$^{-1}$; $V_{eq}=1$ V; $C=10^{-5}$ F/cm$^2$; $R_b=10$ Ω·cm$^2$.

Fig. 3

Simulation for the capacitance measurement by an LCR. All the parameters in simulation are as same as that used in Fig. 2.

Fig. 4

Simulation for the transient current response with the variation of (a) the applied voltage; (b) the accumulation rate. Most the parameters in simulation are as same as that used in Fig. 2 except for $R_b=1$ kΩ·cm$^2$.

**Fig.1**

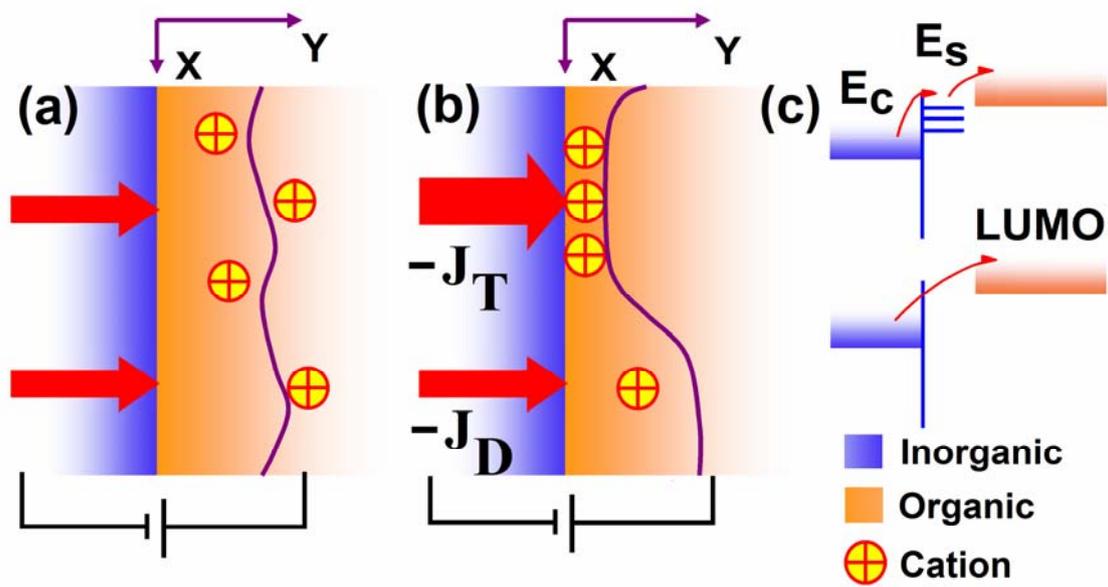

**Fig.2**

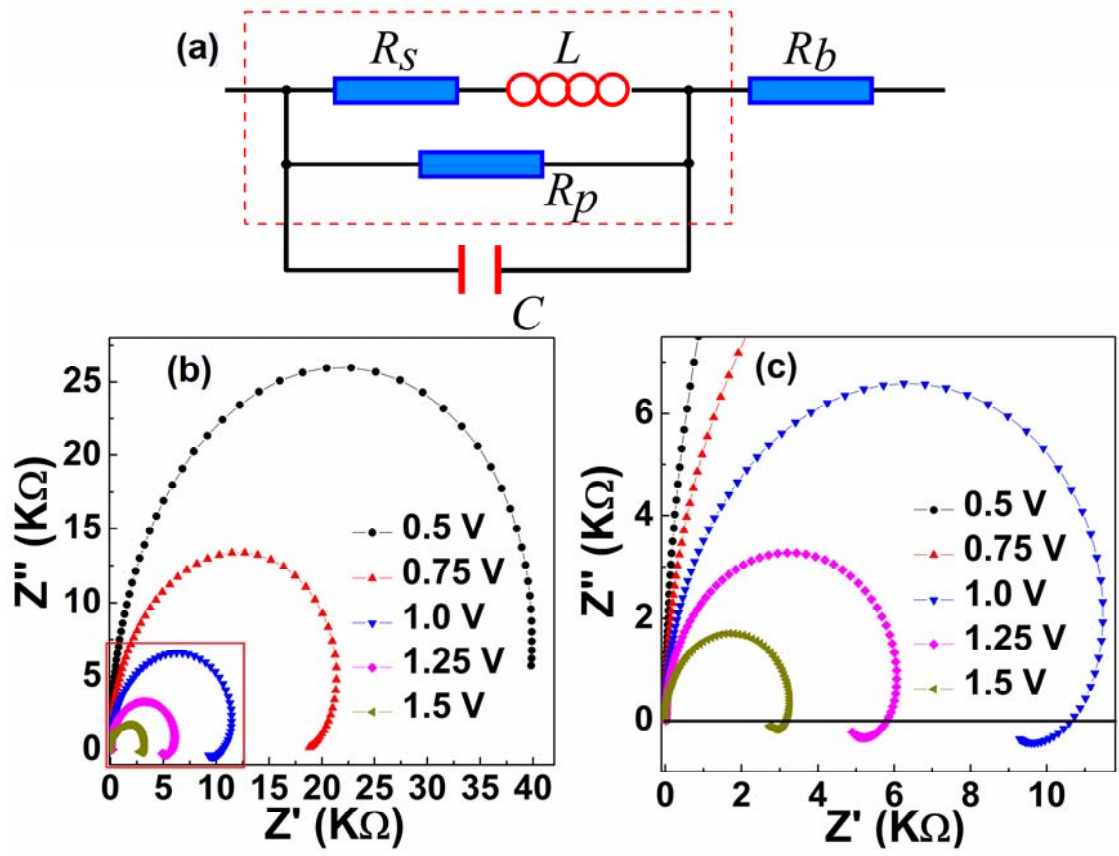

**Fig. 3**

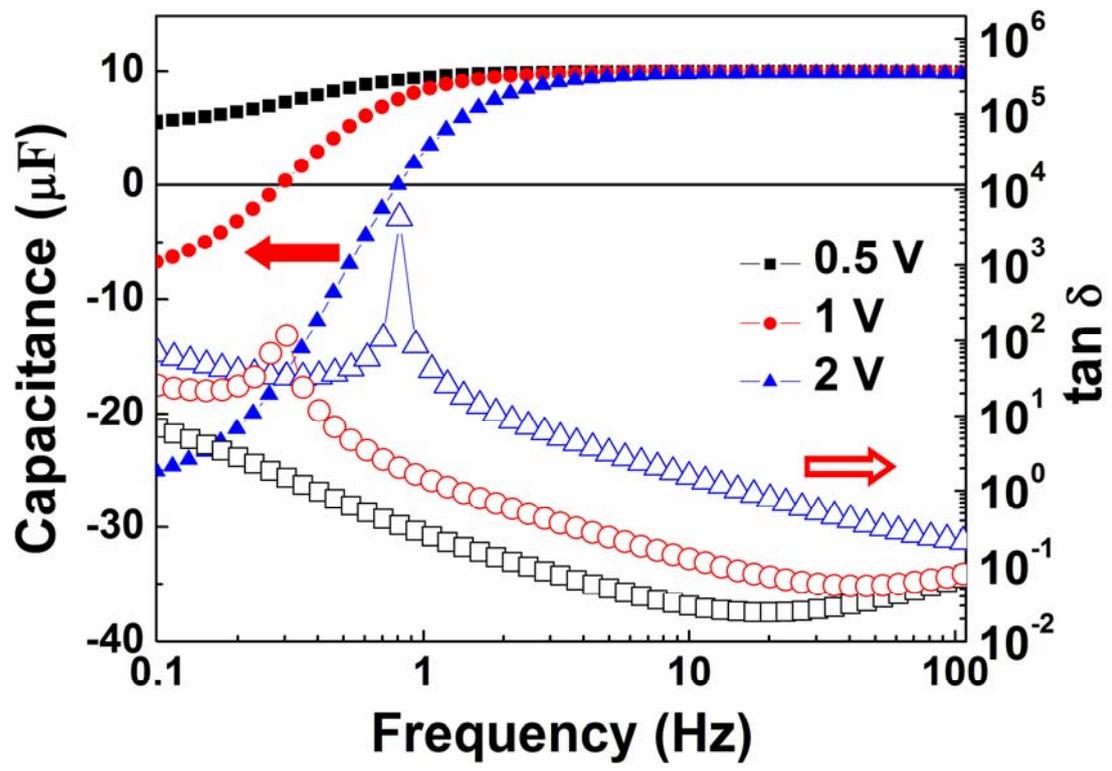

**Fig. 4**

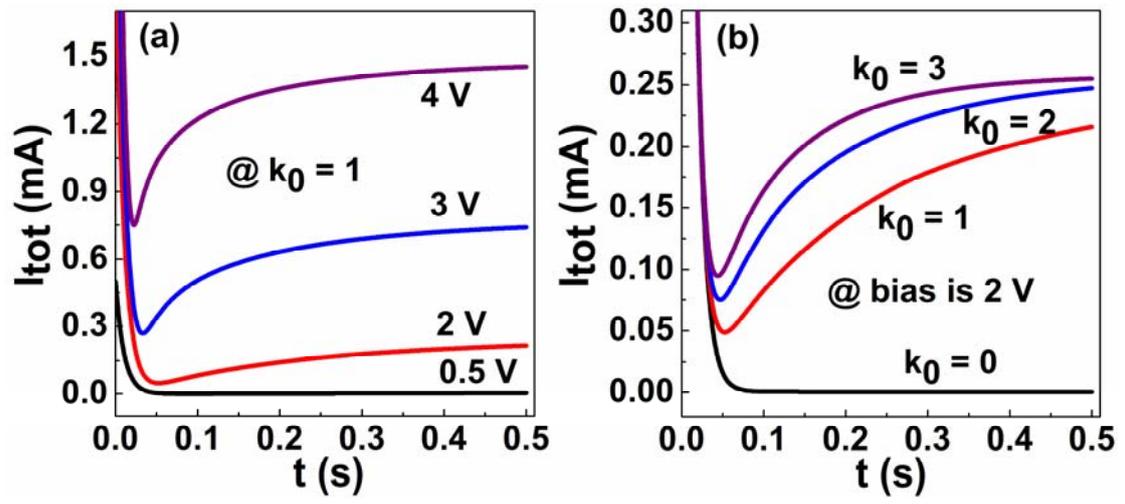